\documentclass[a4paper,10pt,twocolumn]{article}
\usepackage{graphicx}
\usepackage{times}
\usepackage[noline]{algorithm2e}
\usepackage{amsmath}
\usepackage{pdfpages}

\begin{document}
\pagenumbering{gobble}

\title{Multi-Variant Time Constrained FlexRay Static Segment Scheduling}

\author{Jan Dvorak\\
Department of Control Engineering\\
FEE, Czech Technical University in Prague\\Prague, Czech Republic\\
dvoraj57@fel.cvut.cz\\
\and
Zdenek Hanzalek\\
Department of Control Engineering\\
FEE, Czech Technical University in Prague\\Prague, Czech Republic\\
and Porsche Engineering Services, s.r.o.\\
hanzalek@fel.cvut.cz\\
}

\null
\includepdf[pages=1,fitpaper,noautoscale]{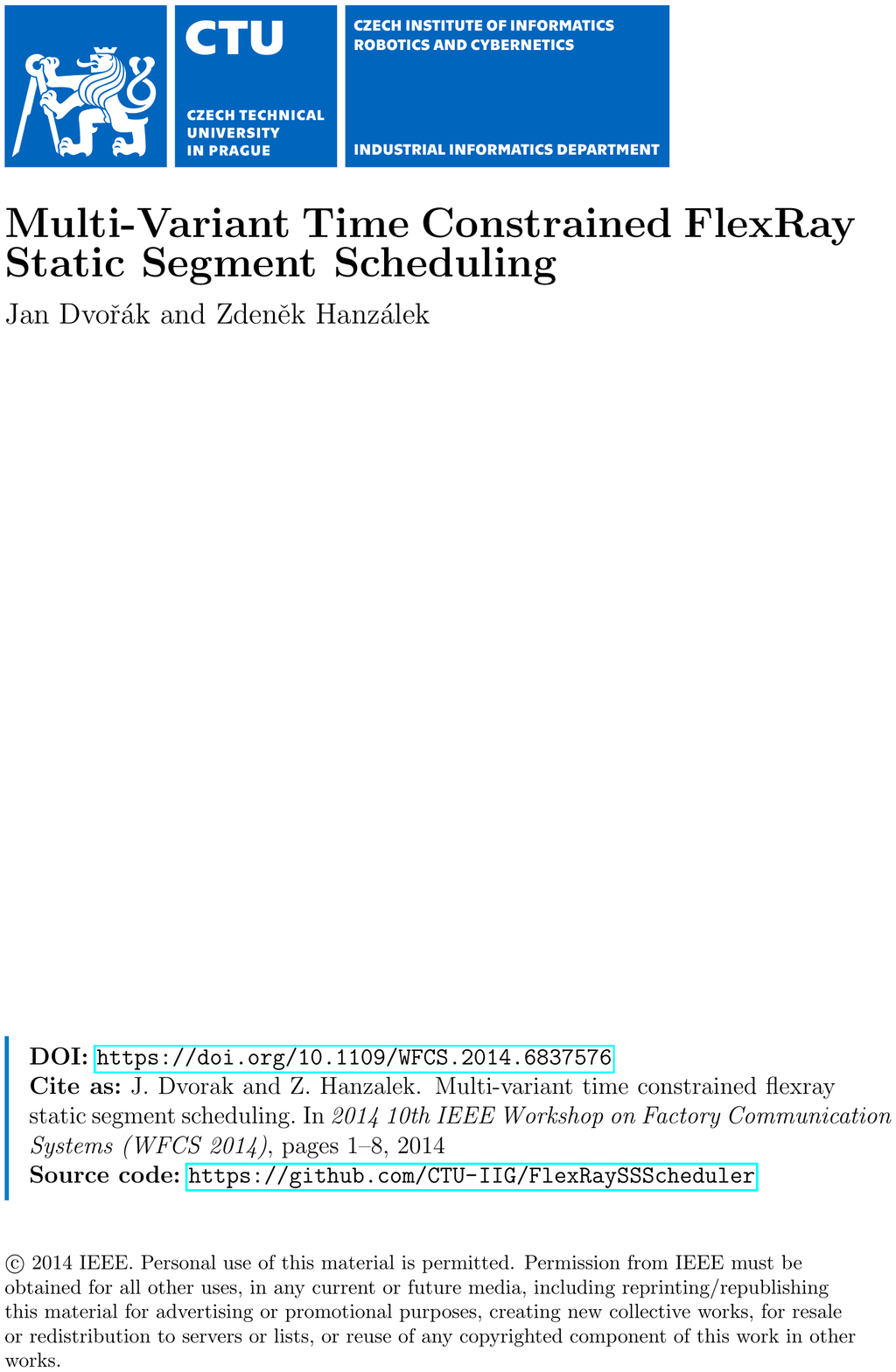}
\date{}
\maketitle

\begin{abstract}
The FlexRay bus is a modern standard used in the automotive industry.
 It offers deterministic message transmission with zero jitter while using time-triggered scheduling in the static segment. When several vehicle variants (i.e. different models and their versions) share the same signal, the car manufacturers require to schedule such signal at the same time in all vehicle variants. This requirement simplifies the signal traceability and diagnostics in different vehicle variants using the same platform and simplifies reuse of components and tools.

In this paper, we propose a first fit based heuristic algorithm which creates the schedules for several vehicle variants at once, while transmitting a given signal at the same time in all the schedules. The scheduling algorithm also takes the time constraints as release dates and deadlines into account. Finally, different algorithm versions are compared on benchmark sets and low computational time demands are validated on large instances.
\end{abstract}

\section{Introduction}

The current automotive industry produces vehicle models containing a lot of electronic control units (ECUs). These units are used to control almost everything from directional indicators to the driver assistance system. Moreover in the upcoming vehicle models x-by-wire systems are replacing mechanic and hydraulic control systems. Consequently, the demand of the bus bandwidth will increase significantly and more criticality-related requirements will need to be satisfied. The FlexRay standard has been designed to handle this situation. The bus offers ten times more bandwidth compared to the CAN bus. A~static segment with time division multiple access can be used for time critical signals. In this segment, signals are transmitted to the bus at exact time points determined by a schedule. The schedule must be known in advance.

\subsection{Motivation}
It is s basic practice for automotive concerns that many vehicle models are built on a common technological platform. For example, the vehicle models as Audi A3, SEAT Leon, Volkswagen Golf and \v{S}koda Octavia share a modular construction of the MQB platform. Moreover, these vehicle models also have many versions (with a basic or advanced ECUs with different sensors etc.). In order to simplify the reuse of components, it is desired to have the inner vehicle communication as similar as possible for all the models and variants. From this perspective, it would be ideal to create just one schedule for all variants with all theirs signals but such a schedule would have a very low utilization of the bus and, consequently, it would limit the number of signals. We need to deal with this and find some clever solution. In our case, we follow the practice where the same signals are placed in the same positions in all schedules they participate and each vehicle variant has a schedule which differs in positions of specific signals only. With this requirement, it is, for example, easier to develop a diagnostic tool because one tool can then be used for all vehicle variants (shortly variants). From another point of view, it also simplifies the configuration of ECUs, typically supplied by third parts, because one bus configuration of the ECU may fit to several variants. In the consequence, the less expensive tests are involved because many mistakes (related to the time dependent electromagnetic interferences for example) are eliminated.

There are already several algorithms that are used for the FlexRay static segment scheduling, but these are suited for one variant schedules only. Our new scenario considers new algorithm that creates the schedules for all the variants at once. We call this problem Multi-Variant scheduling.

\subsection{FlexRay overview}
The FlexRay standard describes a new generation bus developed to satisfy the performance and safety requirements in the advanced automotive industry. It is able to operate with data transmission rates up to 10~Mb/s. The FlaxRay communication protocol was further designed to fit all criticality constraint requirements that arise in automotive industry.

The standard offers two channels to be used for communication. These channels can be used for independent communication as well as for increasing the fault tolerance. We can use different network topologies such as a bus topology or an active star topology. It is also possible to use a hybrid topology and different topologies in both channels. A~set of ECUs connected to the shared FlexRay network is denoted as a cluster. Each ECU connected to this network is called a node. The node can transmit/receive information to/from a bus in predetermined time intervals. A few special nodes (ColdStart nodes and Sync nodes ) are used for starting the bus up and the bus time synchronization. 

\begin{figure}[h]
\resizebox{\columnwidth}{!}
{
\includegraphics{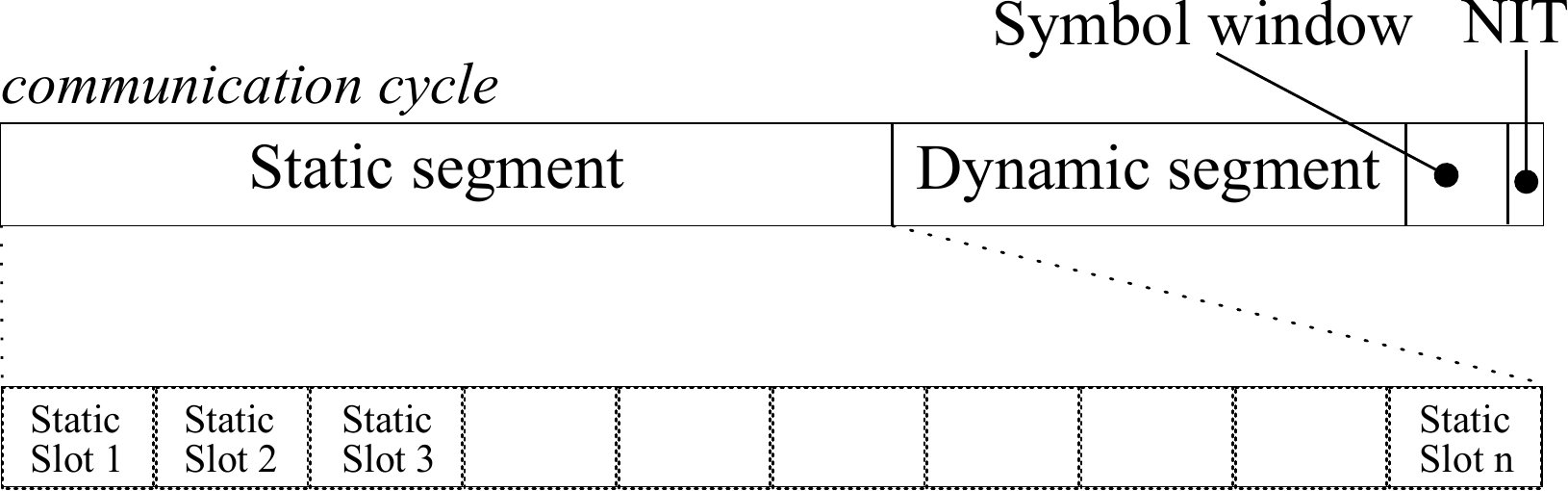}
}
\caption{FlexRay communication scheme}
\label{Fig:CycleScheme}
\end{figure}
The FlexRay communication is organized in cycles. Each communication cycle has its own six-bit cycle number. Thus, there can be up to 64 different cycles. The set of these different cycles is called a hyperperiod and it is \mbox{periodically} repeated. One communication cycle (presented in~Figure~\ref{Fig:CycleScheme}) consists of four segments:
\begin{itemize}
\setlength{\itemsep}{0pt}
\setlength{\parskip}{0pt}
\item Static segment
\item Dynamic segment
\item Symbol window
\item Network idle time
\end{itemize}
Only the static segment and network idle time are mandatory. In the static segment, the time critical signals are exchanged using a time-triggered scheme based on TDMA. The data structure used by the nodes to transmit the data within the static segment is called a frame. The schedule that controls when the frame is transmitted to the network must be known before the bus is started. The dynamic segment fulfills the requirements for event-triggered communication. In this segment, dynamic minislotting based scheme is employed. The symbol window is used for network management messages and special symbol broadcasting. The network idle time (also referred as NIT) is the last segment and during this time no communication takes place. In this time, each node synchronizes its own inner clocks before a new communication cycle starts. 

In this paper, we will focus only on the static segment. The static segment is divided into time intervals of the same duration called static slots. Each frame of the static segment within the hyperperiod is identified by its cycle and slot number. A given slot is reserved for a given node (i.e. the frames transmitted in a given slot need to be from one node in all cycles). The frame can contain more than one signal but the sum of payloads of these signals must not exceed the duration of the slot.

\subsection{Related works}
Several papers were published recently that focus on the FlexRay protocol and particularly the static segment scheduling problem. The FlexRay 3.0.1 is described in detail by the FlexRay\texttrademark Communication System Protocol Specification~\cite{FlexRay}. In the automotive industry, this bus is used together with the AUTOSAR Specification~\cite{AutosarRequirements, AutosarInterface}. Nowadays, BMW or Audi use the FlexRay bus in several series-production vehicles. BMW presented its real-world benefits in~\cite{bmw}.

A~milestone in the static segment scheduling area is the article~\cite{OptimalScheduling} where the transformation of the basic static segment scheduling problem without time constraints to a two-dimensional bin packing problem was introduced. The authors presented an ILP model and also a successful heuristic based on the first fit heuristic for the bin packing problem. The objective is to minimize the number of the scheduled slots and obtain an extensible schedule. The time-constrained problem was proposed in~\cite{TwoStage}. This paper employs an idea of two-stage scheduling, when in the first step, frame packing is performed. After packing the signals to the frames, a frame scheduling algorithm creates the schedule. The response time analyze for the rate monotonic scheduling of the static segment was proposed in ~\cite{OnLineScheduler}. But this paper does not meet the requirements of the AUTOSAR specification and also require modifications of the middleware.

A~genetic algorithm for the FlexRay scheduling is introduced in~\cite{GA1, GA2}. These papers take care not only about the static segment scheduling but also about the placement of tasks into the nodes. A new idea of a switched FlexRay network is presented in~\cite{Switched}. In the switched network, an active star hub is replaced by a switch. It offers the possibility to send different frames in different branches in the same time interval. The branch-and-price based algorithm combined with the branch-and-bound was developed for such frame scheduling. Moreover, the first fit decreasing heuristic was presented which returns good results for input cases even if the exact algorithm does not finish at the appropriate time. An ILP model for scheduling of signals that can have a jitter in the period is proposed in~\cite{Message}. The constraint logic programming formulation (CLP) for fault-tolerant scheduling in the FlexRay static segment is introduced in~\cite{Fault}.

The concept of holistic time analysis of the FlexRay communication protocol is presented in~\cite{Holistic} and~\cite{Time}. The scheduling algorithm and time analysis for the dynamic segment are proposed in~\cite{Dynamic}.

All these articles focus on creating independent schedules and no shared constraints are taken into account. 

In the computer science, a similar problem as multi-variant scheduling is the multiprocessor task scheduling, where some tasks needs two or more processors simultaneously to be executed. The survey for this problem is in~\cite{MultiprocessorSurvey}. In~\cite{MultiprocessorAllocation} the problem complexity is investigated for the multiprocessor task scheduling problem with prespecified processor allocations. It means that it is known in advance exactly which set of processors will be used to execute each task. But in the multiprocessor task scheduling problem each task is executed only once, thus it has no period assigned.

\subsection{Paper outline}
The paper is organized as follows: Section 2 describes the multi-variant time constrained static segment scheduling problem and the signal set used for benchmarking. In Section 3, new efficient data structures are introduced and the first fit based heuristic algorithm is proposed which are the main contributions of this paper. Computational efficiency and performance evaluation of the experiments are presented in Section 4, Section 5 concludes the paper.
\section{Problem statement}

The problem addressed in this paper is to create the FlexRay static segment schedules for a set of variants. The objective is to find an assignment of a signal set $S$ into slots and cycles such that the number of the used slots is minimized (i.e. the length of the longest static segment, of all variants, is the shortest possible). 

The FlexRay network configuration consists of many parameters including the cycle length, the number of static slots in the static segment, the duration of the static slot, duration of the NIT segment, etc. These parameters are usually predefined by the system designer and further followed by the ECU suppliers. We assume that these parameters are known and are not the part of the optimization process. The following setting based on the BMW network design~\cite{dataset} is used in this paper. The communication cycle duration $F$ is equal to~5 ms where the static segment takes 3~ms and the rest of the communication cycle is filled by the dynamic segment, symbol window and NIT. There are 75 static slots each with duration of 0.04~ms. The frame payload $W$ is in the range from 32 to 128~bits. Each signal $i$ has the following parameters defined (the same denotation is used in~\cite{TwoStage}): 
\begin{tabular}{p{0.1cm}p{7.2cm}}
$N_i$&- unique identifier of the ECU, which transmits signal~$i$, \\
$T_i$&- the signal period, the signal is assumed to be transmitted only once in the FlexRay cycle, \\
$O_i$&- the signal release date, it is the latest time after which the first instance of the signal is produced in relation to the start of the hyperperiod, \\
$C_i$&- signal length in bits, \\
$D_i$&- the deadline associated to the signal $i$, it represents the maximum age acceptable at the consumer endpoint. \\
\end{tabular}

Moreover, in the multi-variant scheduling, each signal can be used by one or more variants. For this reason, \mbox{binary} matrix $V_{i,j}$ is introduced as follows: $V_{i,j}$ is equal 1 if the variant $j$ uses the signal $i$ and 0 otherwise. If two or more variants should use signal $i$ in the schedule, this signal must be placed at the same position (cycle, slot, even offset in the frame) in all of these schedules. We call this a shared constraint. This is the reason why it is not possible to create all those schedules independently.

The next assumption is the usage of the AUTOSAR communication stack~\cite{AutosarRequirements}, which is the de facto standard for automotive applications and it is also considered for other industry applications such as aviation. In this context, the AUTOSAR frame contains processing data units (PDU) where each PDU consists of one or more signals. In the AUTOSAR context, it is defined that scheduling is static and cannot be changed during the runtime. It is further assumed that the maximal signal size cannot exceed the data payload of the frame. It means that the signal cannot be split into two or more frames. Moreover, in the AUTOSAR specification, it is defined that one slot is strictly assigned to one node. So it is not possible to use multi-node multiplexing because the slot is reserved for the transmission to the same ECU in every cycle.

\subsection{Signal set}

The scheduling approaches are applied to the modified Society of Automotive Engineers (SAE) benchmark signal set. The SAE report describes a set of signals sent between seven different subsystems in an electric car prototype. The Basic SAE signal set defines 53 sporadic and periodic signals. A~periodic message has a fixed period 5, 10, 100 or 1000~ms, and implicitly requires the latency to be less than or equal to this period. For our purposes, these periods must be resampled because the \mbox{AUTOSAR} specification only allows periods which are equal $\{m\cdot2^n \mid n=1\dots6\}$, where $m$ is the shortest period. 
Sporadic messages have latency requirements imposed by the application: for example, all messages sent as a result of a driver action have a latency requirement of 20~ms. The reader is referred to the work of Kopetz~\cite{benchmark} for a more detailed benchmark description. Due to the high FlexRay bandwidth and electronic equipment's requirements in today’s cars we extended the SEA benchmark using NETCARBENCH~\cite{netcarbench} to increase the number of signals with the same probability distribution function of parameters. The newly created signal sets have up to 3000 signals per node and a few added parameters (such as release date and deadline). The signals are exchanged among up to 23 nodes. Several signals are sent only on a change event where the maximum deadline is defined. These signals are transformed to periodic signals with periods equal to deadlines.

\subsection{Time constraints}
In order to simplify the problem, we consider release date $O_i$ and deadline $D_i$ to be rounded to the start of the earliest and the end of the latest complete cycle respectively in which we can transmit signal $i$ (as in~\cite{TwoStage}). This simplification is adequate since the precise specification of the release dates and deadlines have an influence only if these values fall in the static segment. But if they fall in the dynamic segment, they are rounded to the length of the cycle anyway. This simplifies our scenario because the release dates and deadlines can be only multiples of the cycle length. The position of a signal within the static segment of a particular cycle is insignificant with respect to the position of the signal within the hyperperiod. But a situation can arise when an optimal solution is discarded by this transformation because, due to the rounding, we can cut off a few feasible signal positions from the search space and, thus, we may lose some optimal solutions. But we assume that this situation is rare in real cases.
\begin{table}[t]
\centering
\resizebox{\columnwidth}{!}{%
\begin{tabular}{l|c|r|r|r|r}
Signal&Node&Period&Length&Rel. date&Deadline\\
\hline
A~&1       &5 ms  &8 b    &0 ms       &5 ms\\
B       &1       &10 ms &8 b     &0 ms       &10 ms\\
C       &1       &10 ms &8 b     &0 ms       &10 ms\\
D       &1       &20 ms &8 b     &5 ms       &15 ms\\
E       &1       &20 ms &16 b   &10 ms      &15 ms\\
F       &1       &10 ms &16 b   &5 ms       &10 ms\\
G       &2       &20 ms &8 b     &0 ms       &15 ms\\
H       &3       &20 ms &8 b     &0 ms       &15 ms\\
\end{tabular}}
\caption{Parameters of signals in Example 1}
\label{Tab:SigParams}
\resizebox{0.8\columnwidth}{!}{%
\begin{tabular}{l|c|c|c|c|c|c|c|c}
Variant&A&B&C&D&E&F&G&H\\
\hline
I~&1&1&1&1&0&1&1&0\\
II&0&1&1&0&1&1&0&1\\
\end{tabular}}
\caption{Matrix representing assignment of signals to variants considered in Example 1}
\label{Tab:VMatrix}
\end{table}

\subsection{Example 1: Multi-variant signal set}
\begin{figure}[h]
\centering
\resizebox{\columnwidth}{!}
{
\includegraphics{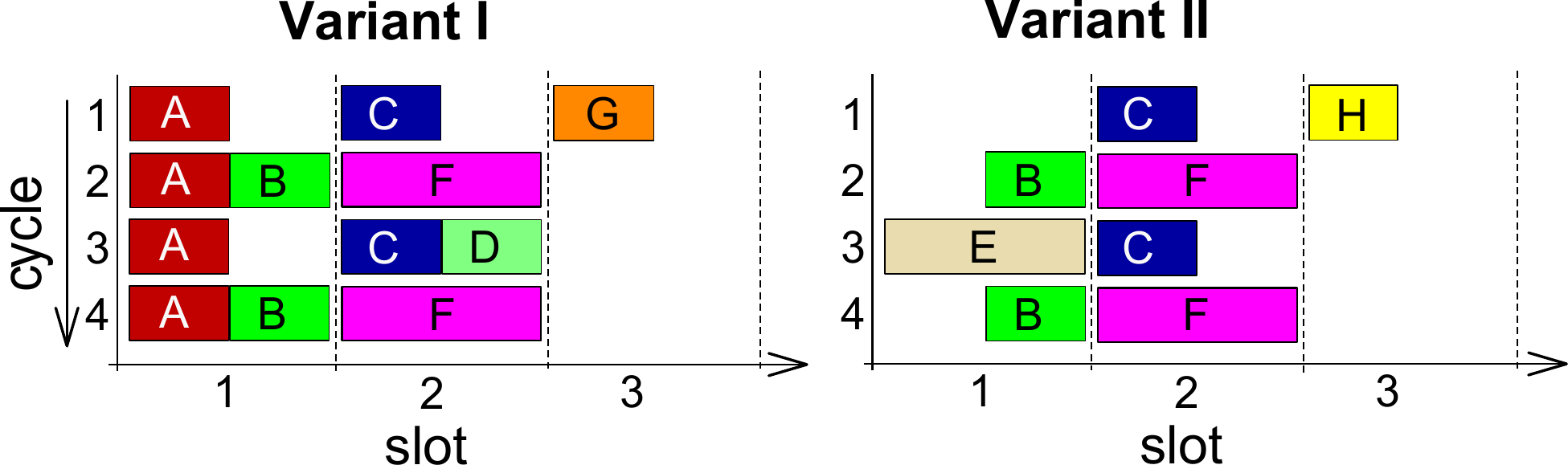}
}
\caption{Feasible schedules for Example 1}
\label{Fig:FeaSchedules}
\end{figure}

For a better understanding we introduce a small example. In our case, the communication cycle duration is set to 5~ms and the frame payload is 16~bits. There are eight signals sent by the three ECUs with the parameters described in Table~\ref{Tab:SigParams}. These signals are to be scheduled to two variants. Matrix $V_{i,j}$ is shown in Table~\ref{Tab:VMatrix}. Variant~I~uses signals $A,B,C,D,F,G$ and variant~II uses signals $B,C,E,F,H$. So it means that signals $B,C,F$ are used in both variants and they have to be scheduled at the same positions. A~feasible (respecting the time constraints and shared constraints) solution for this case is presented in Figure~\ref{Fig:FeaSchedules}. The signal $G$ cannot be placed into the second slot because this slot is used by the first node and the signal $G$ is transmitted by node 2. So a new slot must be allocated for the signal. The same situation is with signal $H$ in variant II. 

\section{Algorithm}
In this section, we will explain the components of the proposed algorithm. But first, let us introduce the main data structures used in our algorithm. 

\subsection{Multischedule}
The most important thing for the algorithm is to use an efficient data structure for the schedule representation. In our case, there are two ways how to do it. In the first, and also most natural way, it would be possible to have one schedule for each variant (like in Figure~\ref{Fig:FeaSchedules}). We call them native schedules. But in this case, during the scheduling process, there is a lot of inefficiency because we have to go through all of these schedules to check the shared constraints. Moreover, if there is a significant number of common signals for several variants, it is also necessary to allocate them in every schedule. But it is sufficient to know their position just in one schedule because this position must be the same for all variants using them. Therefore, in our algorithm, we use a different representation because we assume that many signals are used in all variants.

\begin{figure}[h]
\centering
\resizebox{\columnwidth}{!}
{
\includegraphics{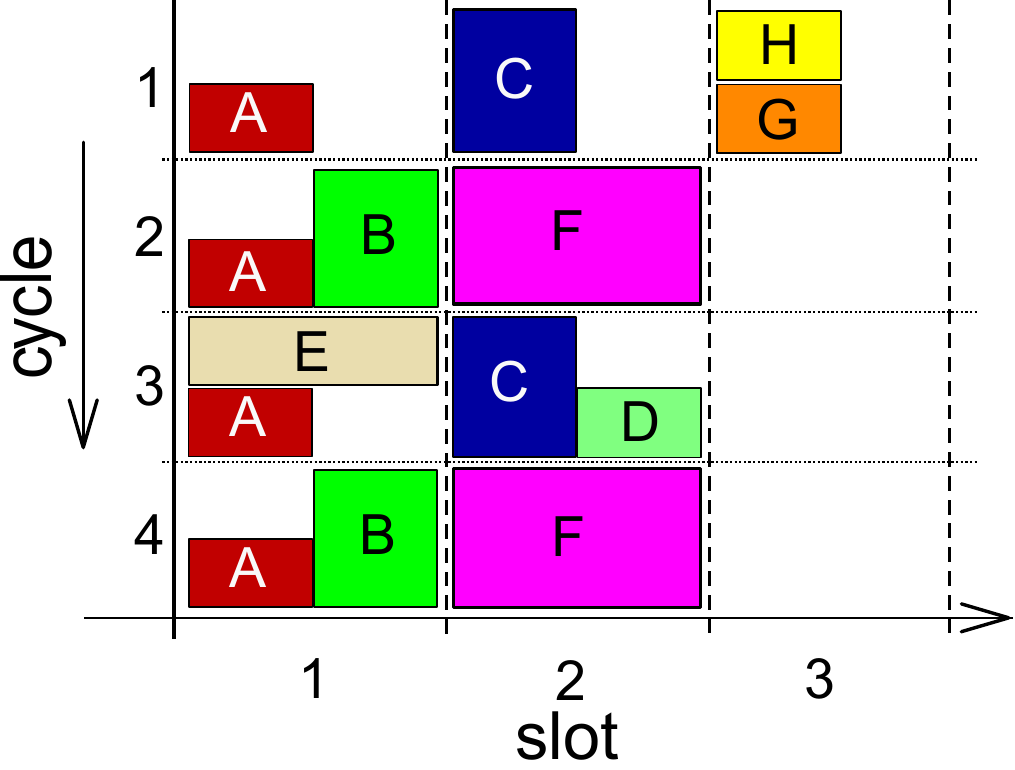}
}
\caption{Feasible multischedule for \mbox{Example 1}}
\label{Fig:Multischedule}
\end{figure}

With this assumption, we can make a more efficient representation. Instead of creating schedules for each variant we create just one shared schedule (multischedule) for all variants. Here, the common signals are placed once and there is no redundancy caused by checking the constraints. But unlike in the native schedule, in the multischedule two or more signals may overlap. Just like the native schedule consists of frames, the multischedule (\textit{MS}) consists of multiframes. We denote these multiframes as $\text{\textit{MS}}_{i,j}$, where $i$ is the cycle number and $j$ is the slot number. The multischedule for Example 1 is presented in Figure~\ref{Fig:Multischedule}. We can see that signal $E$ is scheduled at the same position as signal $A$ in $\text{\textit{MS}}_{3,1}$. It is easy to derive any native schedules from the multischedule by removing all signals that are not used in the particular variant.

\subsection{Mutual exclusion matrices}
For placing signals to the multischedule, we need to know which signals may be mutually overlapped. Overlapping can arise only if these two signals are not scheduled together in any variant. Otherwise, it would result in an infeasible schedule for a variant that uses both signals. Checking if we can overlap two particular signals is a relatively time consuming operation because we need to check all variants if they use both signals or not. But, this information is static and known in advance. Therefore, during the reading of the input instance, we create a Signal Mutual Exclusion Matrix (\textit{SMEM}). \textit{SMEM} is a symmetric binary matrix with the dimension equal to the number of signals. The matrix contains 1 for each pair of signals if these signals are to be scheduled together in some variant or 0 otherwise. 
\begin{table}[h]
\includegraphics[scale=0.43]{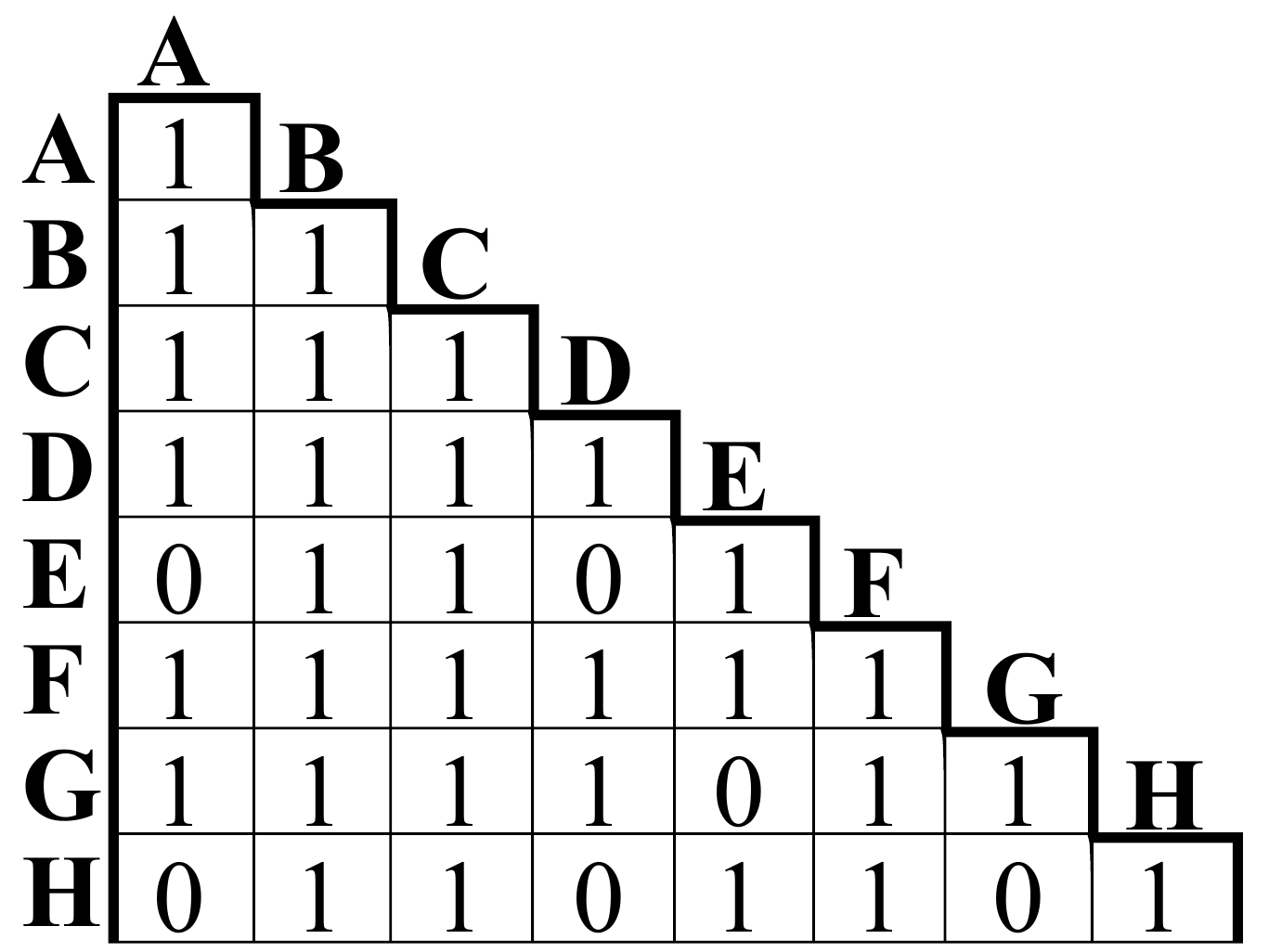}
\caption{Signal mutual exclusion matrix for Example 1}
\label{Tab:SMEM}
\end{table}
Thus two signals $S_1$ and $S_2$ can mutually overlap only if $\text{\textit{SMEM}}_{S1,S2}=0$. In our Example 1, it holds only for pairs $\{A,E\}$; $\{A,H\}$; $\{D,E\}$; $\{D,H\}$; $\{E,G\}$ and $\{G;H\}$ as presented in Table~\ref{Tab:SMEM}.

The multischedule has one extra feature in addition to the native schedule. As can be seen in Figure~\ref{Fig:Multischedule}, one slot in the multischedule can be occupied by more than one node. In our case, in multiframe $\text{\textit{MS}}_{1,3}$ two signals are scheduled. Signal $G$ is from node 2 and signal $H$ is from node 3. To be sure that this behavior does not result in an infeasible schedule, we permit this only if the signals from these two nodes do not appear together in one variant. Thus, we need to know the same information as what holds in \textit{SMEM} for signals, but in this case for nodes. We can also derive this information from the input instance and save it to the matrix. We call this matrix a Node Mutual Exclusion Matrix (\textit{NMEM}). This symmetric binary matrix contains 1 if the nodes are different and appear in some variant together and 0 otherwise. The \textit{NMEM} matrix for our Example~1 is shown in Table~\ref{Tab:NMEM}.

\begin{table}[h]
\includegraphics[scale=0.43]{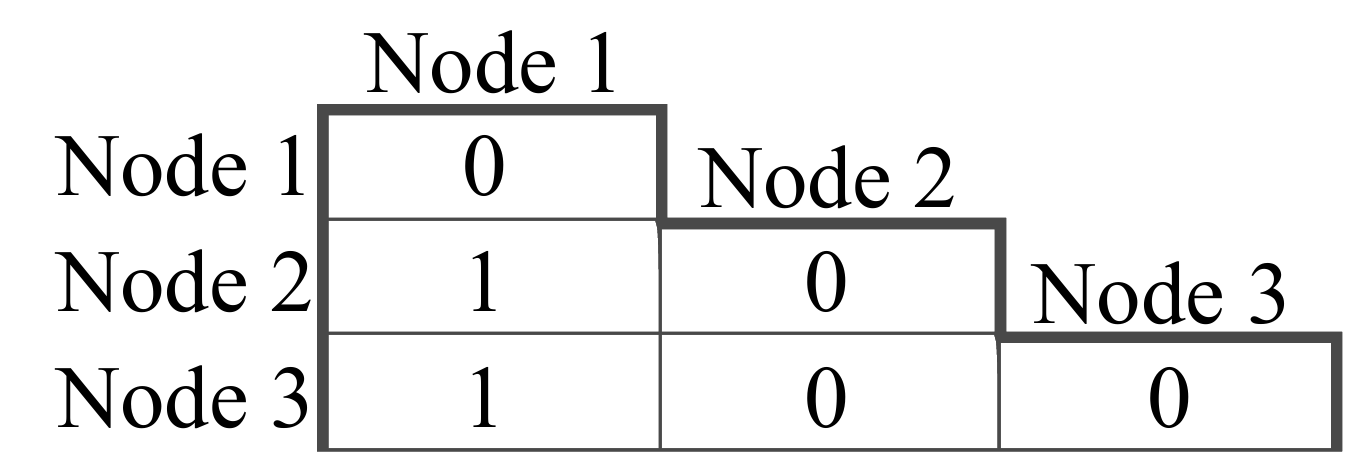}
\caption{Node mutual exclusion matrix for Example 1}
\label{Tab:NMEM}
\end{table}

\subsection{First fit heuristic}
\begin{algorithm}[t]
\SetFuncSty{textsc}
\SetKwFunction{KwPlace}{PlaceSignalToSchedule}
\SetKwFunction{KwSort}{Sort}
\SetKwFunction{KwCalc}{CalculateMEMs}
 \KwIn{$S$}
 (\textit{SMEM}, \textit{NMEM}) $\leftarrow$ \KwCalc{$S$, $V_{i,j}$}\;
 \textit{SL} $\leftarrow$ \KwSort{$S$}\;
initialize \textit{MS}\;
 \For{$i \leftarrow 1$ \KwTo $|\text{\textit{SL}}|$}{
  \KwPlace{\textit{MS}, $\text{\textit{SL}}_i$, \textit{SMEM}, \textit{NMEM}}\;
 }
\caption{First fit algorithm}
\label{Alg:FirstFit}
\end{algorithm} 
In article~\cite{OptimalScheduling}, it was shown that the scheduling problem for the static segment of the FlexRay without time constraints is reducible to the two-dimensional bin packing problem. For this problem, the first fit decreasing heuristic (items are sorted according to decreasing size and then placed one by one to bins in the first suitable position) is known and applicable with an approximation factor 2 (it is proved that solution found by the heuristic can not be worse than twice w.r.t. optimum). But, after adding time constraints, this approximation factor does not hold. So we propose the general first fit heuristic that is initialized by ordered list of signals \textit{SL}.
Used First fit algorithm is outlined in Algorithm~\ref{Alg:FirstFit}.

\begin{algorithm}[h]
\SetFuncSty{textsc}
\SetAlgoLined
\SetKw{KwFn}{PlaceSignalToSchedule}
\SetKwFunction{KwFind}{FindPositionForSignal}
\SetKwFunction{KwPlace}{PlaceSignal}
\SetKwFunction{KwCreate}{AllocateNewSlot}
\SetKw{KwBreak}{break}
 \KwFn{(MS,~signal,~SMEM,~NMEM)}\\
\{ \\
\textit{infeasiblePosition} $\leftarrow true$\;
\While{\textit{infeasiblePosition} $= true$}{
	\textit{placePosition} $= $ \KwFind{MS,~signal,~\\SMEM,~NMEM}\;
	\If{placePosition not found}
	{
	  \KwBreak\;
	}
	(\textit{cycle}, \textit{slot}, \textit{offset})$ \leftarrow$ \textit{placePosition}\;
	\textit{infeasiblePosition} $\leftarrow false$\;
	\While{cycle $<$ hyperperiod}
	{
	   \If{(cycle, slot, offset) is not suitable for signal}
	   {
	      \textit{infeasiblePosition} $\leftarrow true$\;
	      \KwBreak\;
	   }
	   \textit{cycle} += $\text{\textit{T}}_{\text{\textit{signal}}}$\;
	}
 }
\If{infeasiblePosition}
{
   \textit{placePosition} $\leftarrow $\KwCreate{MS}\;
}
\KwPlace{signal, MS, placePosition}\;
\caption{PlaceSignalToSchedule procedure}
\label{Alg:PlaceSignalToSchedule}
\} \\
\end{algorithm}

At the beginning of algorithm, the mutual exclusion matrices are calculated (from signal set $S$ and matrix $V_{i,j}$), then the \textit{SL} list is constructed by ordering the signals from $S$. Then, we allocate the memory and initialize the multischedule \textit{MS}. After that, the signals are sequentially read from \textit{SL} and we place them to the multischedule by the first fit policy.

The most important part of the first fit algorithm is the \mbox{PlaceSignalToSchedule} procedure (Algorithm~\ref{Alg:PlaceSignalToSchedule}). This procedure is responsible for finding a sufficient position (a position where all the constraints are satisfied) and placing the signal to the constructed \textit{MS}. First, the procedure needs to find a feasible position for the first instance of the signal. We denote these instances as the signal job. The FindPositionForSignal procedure finds this position and saves it to the \textit{placePosition} variable. For the signal placement we have to know the cycle number, the slot number and also offset in a frame. It is not possible to deduce the offset in the frame after the scheduling phase as in~\cite{OptimalScheduling}. Because of the shared constraints, it does not hold that if there are enough free bits for the signal in the multiframe we can also find enough free bits in the sequence for placement.

If no feasible position for the first signal job is found then we can finish searching because there is no appropriate position in the already allocated slots. Therefore, we need to allocate a new slot in \textit{MS} and obtain the first feasible position for the signal in this slot and place it there.

\begin{algorithm}[h]
\SetAlgoLined
\SetFuncSty{textsc}
\SetKwFunction{KwFind}{FindSuitableOffset}
\SetKwFunction{KwNodes}{getNodes}
\SetKwFunction{KwNode}{getNode}
\SetKw{KwFn}{FindPositionForSignal}
\SetKw{KwIn}{in}
 \KwFn{(MS, signal, SMEM, NMEM)}\\
\{ \\
(\textit{slot}, \textit{cycle}, $\textit{offset}$) $\leftarrow lastPosition$\;
\ForEach{\textit{slot} after \textit{lastPosition}}
{
\ForEach{node \KwIn \KwNodes{slot}}
{
   \textit{signalNode} = $N_{\text{signal}}$\;
   \If{$\text{\textit{NMEM}}_{\text{\textit{signalNode}}, \text{\textit{node}}}$}
   {
      continue with next slot\;
   }
}
  \ForEach{cycle after lastPosition $< \text{\textit{D}}_{\text{\textit{signal}}}$}
  {
     \textit{offset} = \KwFind{$\text{\textit{MS}}_{\text{\textit{cycle}}, \text{\textit{slot}}}$,\\signal,~\textit{SMEM},~\textit{NMEM}}\;
     \If{offset is found}
     {
	\textit{lastPosition} $\leftarrow$ (\textit{slot}, \textit{cycle}, \textit{offset})\;
	\Return{lastPosition}\;
     }
  }
}
\Return{not found}\;
\} \\
\caption{FindPositionForSignal procedure}
\label{Alg:FindPositionForSignal}
\end{algorithm}

If a feasible position is found by the \mbox{FindPositionForSignal} procedure it does not necessary mean that we can place the signal there. This holds only if \textit{SL} is sorted in an increasing order of periods but not in a general case. Thus, we need to check the conflict for each signal job. If the position of the first signal job is determined, then the positions of all the signal jobs are easy to deduce because of the strict period. Its slot number and offset in the frame are the same, only the cycle number is greater by a multiple of the signal period. If there are free positions for all the signal jobs then the feasible signal position is found. We can stop searching and place the signal to the determined position. Otherwise in the case when some signal job cannot be assigned to the corresponding position in the multiframe we need to find a new position for the first signal job and check its feasibility again. 

The most computationally intensive and complex part is calling the \mbox{FindPositionForSignal} procedure described in Algorithm~\ref{Alg:FindPositionForSignal}. The procedure can be called many times for a single signal. First, we restore the position found by last procedure call because we want to find the next one. If the procedure is invoked for the first time the \textit{lastPosition} variable is equal to the signal release date. Then, we check the multiframes after this position that satisfy the time constraints until some sufficient position is found. But we only need to check the multiframes in the slots in which the transmitting node can operate. Because more than one node can be assigned to one slot in \textit{MS} we iterate through all these nodes and check if there is some conflict. If not, we look for a sufficient multiframe otherwise we can skip this slot. A~multiframe is sufficient if we can find an offset where the first signal job could be placed. This is done by the FindSuitableOffset procedure presented in Algorithm~\ref{Alg:FindSuitableOffset}. 

\begin{algorithm}[h]
\SetFuncSty{textsc}
\SetAlgoLined
\SetKwFunction{KwFind}{FindFirstSuitableOffset}
\SetKwFunction{KwSignals}{getSignals}
\SetKwFunction{KwOffset}{getOffset}
\SetKwFunction{KwAlloc}{allocateMemory}
\SetKw{KwIn}{in}
\SetKw{KwFn}{FindSuitableOffset}
\small
 \KwFn{(frame,~signal,~\textit{SMEM},~\textit{NMEM})}\\
\{ \\
\textit{freeBits} = \KwAlloc{$W$}\;
\ForEach{fs \KwIn \KwSignals{frame}}
{
   \textit{fsOffset} = \KwOffset{fs}\;
   \If{$\text{\textit{SMEM}}_{fs, \text{\textit{signal}}}$}
   {
      \For{$i \leftarrow$ fsOffset \KwTo fsOffset $ + \text{\textit{C}}_{fs}$}
          {
	   \textit{freeBits[i]} $\leftarrow 1$\;
          }
   }
}

\Return{\KwFind{\textit{freeBits[i]}}}\;
\} \\
\caption{FindSuitableOffset}
\label{Alg:FindSuitableOffset}
\end{algorithm}

For placing the signal job to the multiframe, there must be at least as many free bits in the sequence as is in the payload of the signal. The free bit, in our case, means that, in this bit of the multiframe, no signal is placed that could cause a conflict with the currently assigned one. Hence, we create the binary array called \textit{freeBits} that has the same length as is the multiframe payload. Initially, the \textit{freeBits} array contains only zeroes. Then, we iterate through all signals \textit{fs} already assigned to the multiframe. If \textit{fs} is in conflict with the assigned signal all positions in the \textit{freeBits} corresponding to \textit{fs} are set to 1. Finally, each position of the array contains 0 if the bit in the mutiframe is free and 1 otherwise. The \mbox{FindFirstSuitableOffset} function then finds the first position in the \textit{freeBits} array where the sequence of zeroes of length at least equal to the payload of the signal starts. This position is also equal to the first position where the signal can be placed in the multiframe. Thus, this position is returned as a sufficient offset.

\section{Experimental results}
\begin{table}[t]
\centering
\resizebox{\columnwidth}{!}{%
\begin{tabular}{l|r|r|r|r|r|r}
&TS&FF&FFP&FFW&FFL&FFC\\
\hline
$set1$&20.2&30.9&\textbf{19.0}&\textbf{19.0}&41.4&\textbf{19.0}\\
$set2$&20.4&32.5&\textbf{19.2}&\textbf{19.2}&41.6&\textbf{19.2}\\
$set3$&20.4&32.6&\textbf{19.3}&19.4&41.0&\textbf{19.3}\\
$set4$&19.7&29.8&\textbf{18.6}&\textbf{18.6}&43.3&\textbf{18.6}\\
$set5$&11.8&28.8&11.3&11.3&38.2&\textbf{11.1}\\
$set6$&26.4&55.5&24.6&24.6&63.3&\textbf{24.4}\\
$set7$&28.9&47.5&\textbf{28.6}&\textbf{28.6}&49.0&\textbf{28.6}\\
{\small$1ECU500$}&17.6&25.9&17.0&17.0&27.7&\textbf{16.9}\\
{\small$1ECU1000$}&34.3&58.1&33.8&\textbf{33.7}&63.7&\textbf{33.7}\\
{\small$1ECU3000$}&23.2&120.0&22.5&22.6&134.6&\textbf{22.4} \\
\hline \hline
Average&22.3&46.2&21.39&21.4&54.4&\textbf{21.3}\\
\end{tabular}}
\caption{Number of the slots allocated by \mbox{different} algorithms}
\label{Tab:QualityResults}
\end{table}
The proposed algorithm was coded in C++ and tested on a PC with an Intel\textregistered Core\texttrademark2 Duo CPU  (2.8~GHz) and 8~GB RAM memory. For the experiments, the SAE benchmark signal set was used. From this basic set we created seven extended benchmark sets as in~\cite{TwoStage}. Detailed description of the extended sets is as follows:\\
\begin{tabular}{@{}p{0.3cm}p{7.34cm}}
\textit{set1}&- neither release dates nor deadlines are defined\\
&- 3 nodes\\
\end{tabular}
\begin{tabular}{@{}p{0.3cm}p{7.34cm}}
\textit{set2}&- deadlines are not set, release dates are spread across the first five communication cycles\\
&- 3 nodes\\
\end{tabular}
\begin{tabular}{@{}p{0.3cm}p{7.34cm}}
\textit{set3}&- each deadline is set in the last third of the signal period, each release date is set in the first five communication cycles \\
&- 3 nodes\\
\end{tabular}
\begin{tabular}{@{}p{0.3cm}p{7.34cm}}
\textit{set4}&- similar to \textit{set2} \\
&- 3 nodes\\
\end{tabular}
\begin{tabular}{@{}p{0.3cm}p{7.34cm}}
\textit{set5}&- deadlines are not set, release dates are spread across the first five communication cycles \\
&- 6 nodes\\
&- 64-bit payload\\
\end{tabular}
\begin{tabular}{@{}p{0.3cm}p{7.34cm}}
\textit{set6}&- similar constraints as \textit{set5}, but with a higher number of signals \\
&- 6 nodes\\
\end{tabular}
\begin{tabular}{@{}p{0.3cm}p{7.34cm}}
\textit{set7}&- neither release dates nor deadlines are defined, but there are around 40 signals per node in average \\
&- 23 nodes\\
\end{tabular}
All of these sets contain 500 to 1000 signals. Because each of them was scheduled in less then 0.1~s, three extra benchmark sets were prepared to measure the computation time. All signals are transmitted from one node. This results in more computational cost because we cannot skip any already allocated slot (\textit{NMEM} is equal 0). The description of these extra sets is as follows:\\
\begin{tabular}{@{}p{1.5cm}p{6.14cm}}
\textit{1ECU500}&- deadlines are randomly defined, each release date is set in the first five communication cycles \\
&- about 500 signals\\
\end{tabular}
\begin{tabular}{@{}p{1.5cm}p{6.14cm}}
\textit{1ECU1000}&- the same as \textit{1ECU500}\\
&- about 1000 signals\\
\end{tabular}
\begin{tabular}{@{}p{1.5cm}p{6.14cm}}
\textit{1ECU3000}&- the same as \textit{1ECU500}\\
&- 128-bit payload\\
&- about 3000 signals\\
\end{tabular}
The described datasets have a payload set to 32 bits with the exception of \textit{set5} and \textit{1ECU3000}. Each set consists of 10 test cases. In all of these test cases, twenty variants were randomly generated such that all signals were used. Each variant had a random number from 0 to 0.7 assigned. The signals are then added to the variant with the probability denoted by this number. So we are simulating different vehicle classes (from the cheap variant with only a few signals to the expensive one with many signals).

\begin{table}[t]
\centering
\resizebox{\columnwidth}{!}{%
\begin{tabular}{l|r|r|r|r|r|r}
&TS[s]&FF[s]&FFP[s]&FFW[s]&FFL[s]&FFC[s]\\
\hline
1ECU500&0.202&0.032&0.059&0.060&0.039&0.068\\
1ECU1000&0.877&0.168&0.243&0.248&0.202&0.262\\
1ECU3000&16.09&1.314&1.099&1.149&1.950&1.141\\
\end{tabular}}
\caption{Runtime of the algorithms}
\label{Tab:PerformanceResults}
\end{table}

We tested several algorithms for comparison. The TS algorithm is the modified two-stage algorithm described in~\cite{TwoStage} for the time constrained static segment scheduling. It packs the signals into multiframes first and then it schedules the already assembled multiframes. The remaining algorithms are using the proposed first fit algorithm with different ordering. The FF (FirstFit) algorithm uses no ordering. It just schedules the signals in the order in which they are read from the input instance. The FFP (FirstFitPeriod) algorithm sorts signals according to the signal period in an increasing order. The FFW (FirstFitWindow) algorithm uses an increasing order of time windows (the gaps between the release dates and deadlines) so the signals that only have a few cycles where they can be placed are scheduled first. The FFL (FirstFitLength) algorithm schedules signals in a decreasing order of the signal payload. The FFC (FirstFitCombined) algorithm is a combination of multiple ordering. In this case, the signals are sorted by the stable sorting algorithm in sequence according to: decreasing payload, increasing windows, increasing period and increasing node number. 

The experimental results are presented in Table~\ref{Tab:QualityResults}. In this table, the benchmark sets are situated in the rows and the different algorithms are in the columns. Each cell contains an average number of the used slots computed from all test cases of the benchmark set. In the last row there is the overall average for the particular algorithm.

Table~\ref{Tab:PerformanceResults} presents the runtime of the algorithms on the extra benchmark sets. The table is organized in the same way as Table~\ref{Tab:QualityResults}. As we can see, FFP, FFW and FFC return good results in terms of the quality and also in better time than FFL and FF. The TS algorithm returns only a little bit worse solutions but it consumes much more time for the test cases with a greater number of signals per node. The FFL algorithm returns worse results than FF. It is due to the nature of the benchmark set where signals with bigger payloads often have a bigger period too. If the signals with a big period are placed first then it is also more complicated to find a feasible position for the other signals. The consequence of this behavior is also a bad performance.

\section{Conclusion}
In this paper, we described the heuristic based on first fit policy for solving a multi-variant time constrained static segment scheduling problem. We introduced new data structures for better efficiency and performance and we showed that the idea of the first fit decreasing algorithm from the bin packing problems area is useful for solving a multi-variant scheduling problem.

We tested our proposed algorithm with several types of signal ordering. The best results were obtained by the combination of multiple ordering. But the most important outcome is a good quality result with ordering according to the increasing period. There is possibility to optimize the FFP algorithm to gain better performance.  It is also possible, because of relatively low computational complexity, to use the proposed algorithm as a basis for a genetic algorithm or other metaheuristic. In the future, the method for incremental multi-variant scheduling should be proposed.

\section*{Acknowledgment}
This work was supported by the Grant Agency of the Czech Republic under the Project GACR P103/12/1994.

\bibliographystyle{ieeetr}
\bibliography{ms}

\begin{thebibliography}{10}

\bibitem{FlexRay}
{FlexRay consortium}, ``Flexray communications system protocol specification
  version 3.0.1,'' Oct 2010.

\bibitem{AutosarRequirements}
{AUTOSAR Development Partnership}, ``Autosar requirements on flexray v4.0.1,''
  Oct 2013.

\bibitem{AutosarInterface}
{AUTOSAR Development Partnership}, ``Autosar specification of flexray interface
  v3.5.0,'' Oct 2013.

\bibitem{bmw}
BMW, ``Goals and architecture of flexray at bmw,'' Mar 2007.

\bibitem{OptimalScheduling}
M.~Lukasiewycz, M.~Gla\ss, J.~Teich, and P.~Milbredt, ``Flexray schedule
  optimization of the static segment,'' in {\em Proceedings of the 7th IEEE/ACM
  International Conference on Hardware/Software Codesign and System Synthesis},
  CODES+ISSS '09, (New York, NY, USA), pp.~363--372, ACM, 2009.

\bibitem{TwoStage}
{Z. Hanzalek, D. Benes and D. Waraus}, ``Time constrained flexray static
  segment scheduling,'' in {\em preprints of the 10th International Workshop on
  Real-Time Networks (RTN’2011), In conjunction with ECRTS 2011, Porto,
  Portugal, July 5th}, 2011.

\bibitem{OnLineScheduler}
R.~Lange, F.~Vasques, P.~Portugal, and R.~de~Oliveira, ``Guaranteeing real-time
  message deadlines in the flexray static segment using a on-line scheduling
  approach,'' in {\em Factory Communication Systems (WFCS), 2012 9th IEEE
  International Workshop on}, pp.~301--310, May 2012.

\bibitem{GA1}
S.~Ding, H.~Tomiyama, and H.~Takada, ``An effective ga-based scheduling
  algorithm for flexray systems,'' {\em IEICE - Trans. Inf. Syst.}, vol.~E91-D,
  pp.~2115--2123, Aug. 2008.

\bibitem{GA2}
S.~Ding, ``Scheduling approach for static segment using hybrid genetic
  algorithm in flexray systems,'' in {\em Computer and Information Technology
  (CIT), 2010 IEEE 10th International Conference on}, pp.~2355--2360, 2010.

\bibitem{Switched}
T.~Schenkelaars, B.~Vermeulen, and K.~Goossens, ``Optimal scheduling of
  switched flexray networks.,'' in {\em DATE}, pp.~926--931, IEEE, 2011.

\bibitem{Message}
K.~Schmidt and E.~Schmidt, ``Message scheduling for the flexray protocol: The
  static segment,'' {\em Vehicular Technology, IEEE Transactions on}, vol.~58,
  no.~5, pp.~2170--2179, 2009.

\bibitem{Fault}
B.~Tanasa, U.~Bordoloi, P.~Eles, and Z.~Peng, ``Scheduling for fault-tolerant
  communication on the static segment of flexray,'' in {\em Real-Time Systems
  Symposium (RTSS), 2010 IEEE 31st}, pp.~385--394, 2010.

\bibitem{Holistic}
Y.~Hua, X.~Liu, and W.~He, ``Hosa: Holistic scheduling and analysis for
  scalable fault-tolerant flexray design,'' in {\em INFOCOM, 2012 Proceedings
  IEEE}, pp.~1233--1241, 2012.

\bibitem{Time}
T.~Pop, P.~Pop, P.~Eles, Z.~Peng, and A.~Andrei, ``Timing analysis of the
  flexray communication protocol,'' in {\em Real-Time Systems, 2006. 18th
  Euromicro Conference on}, pp.~11 pp.--216, 2006.

\bibitem{Dynamic}
K.~Schmidt and E.~Schmidt, ``Schedulability analysis and message schedule
  computation for the dynamic segment of flexray,'' in {\em Vehicular
  Technology Conference Fall (VTC 2010-Fall), 2010 IEEE 72nd}, pp.~1--5, 2010.

\bibitem{MultiprocessorSurvey}
M.~Drozdowski, ``Scheduling multiprocessor tasks -- an overview,'' 1996.

\bibitem{MultiprocessorAllocation}
J.~Hoogeveen, S.~van~de Velde, and B.~Veltman, ``Complexity of scheduling
  multiprocessor tasks with prespecified processor allocations,'' {\em Discrete
  Applied Mathematics}, vol.~55, no.~3, pp.~259 -- 272, 1994.

\bibitem{dataset}
M.~P. Josef~Berwanger and A.~Schedl, {\em FlexRay startet durch - FlexRay-
  Bordnetz fr Fahrdynamik und Fahrerassistenzsysteme}.
\newblock 2008.

\bibitem{benchmark}
H.~Kopetz, ``A solution to an automotive control system benchmark,'' in {\em
  Real-Time Systems Symposium, Proceedings.}, pp.~154--158, 1994.

\bibitem{netcarbench}
N.~N. Christelle~Braun, Lionel~Havet, ``Netcarbench: A benchmark for techniques
  and tools used in the design of automotive communication systems.,'' in {\em
  proceedings of the 7th IFAC International Conference on Fieldbuses \&
  Networks in Industrial \& Embedded Systems (FeT 2007), Toulouse, France}, Nov
  2007.

\end{thebibliography}
\end{document}